# A NEW CRITERION FOR BAR–FORMING INSTABILITY IN RAPIDLY ROTATING GASEOUS AND STELLAR SYSTEMS. I. AXISYMMETRIC FORM

Dimitris M. Christodoulou[1], Isaac Shlosman[2,3], and Joel E. Tohline[4]

November 7, 1994

## ABSTRACT

We analyze previous results on the stability of uniformly and differentially rotating, self–gravitating, gaseous and stellar, axisymmetric systems to derive a new stability criterion for the appearance of toroidal, $m = 2$ Intermediate or I–modes and bar modes. In the process, we demonstrate that the bar modes in stellar systems and the $m = 2$ I–modes in gaseous systems have many common physical characteristics and only one substantial difference: because of the anisotropy of the stress tensor, dynamical instability sets in at lower rotation in stellar systems. This difference is reflected also in the new stability criterion.

The new stability parameter $\alpha = T_J/|W|$ is formulated first for uniformly rotating systems and is based on the angular momentum content rather than on the energy content of a system. ($T_J \equiv L\Omega_J/2$, $L$ is the total angular momentum, $\Omega_J$ is the Jeans frequency introduced by self–gravity, and $W$ is the total gravitational potential energy.) For stability of stellar systems $\alpha \leq 0.254$–$0.258$ while $\alpha \leq 0.341$–$0.354$ for stability of gaseous systems.

For uniform rotation, one can write $\alpha = (ft/2)^{1/2}$, where $t \equiv T/|W|$, $T$ is the total kinetic energy due to rotation, and $f$ is a function characteristic of the topology/connectedness and the geometric shape of a system. Equivalently, $\alpha = t/\chi$, where $\chi \equiv \Omega/\Omega_J$ and $\Omega$ is the rotation frequency. Using these forms, $\alpha$ can be extended to and calculated for a variety of *differentially* rotating, gaseous and stellar, axisymmetric disk and spheroidal models whose equilibrium structures and stability characteristics are known. In this paper, we also estimate $\alpha$ for gaseous toroidal models and for stellar disk systems embedded in an inert or responsive "halo." We find that the new stability criterion holds equally well for all these previously published axisymmetric models.

*Subject headings:*  galaxies: evolution – galaxies: structure – hydrodynamics – instabilities – stars: formation

[1] Virginia Institute for Theoretical Astronomy, Department of Astronomy, University of Virginia, P.O. Box 3818, Charlottesville, VA 22903

[2] Department of Physics & Astronomy, University of Kentucky, Lexington, KY 40506

[3] Gauss Foundation Fellow

[4] Department of Physics & Astronomy, Louisiana State University, Baton Rouge, LA 70803

1   INTRODUCTION

Many classical investigations performed over the past four centuries were concerned with the equilibrium structures and stability properties of homogeneous, rotating, self–gravitating, fluid masses. Chandrasekhar (1969) elegantly summarized and brought in compact form all these classical studies in his monograph "Ellipsoidal Figures of Equilibrium" (hereafter referred to as EFE). Besides their mathematical tractability, such fluid models were thought to be applicable to the structure and evolution of stars and planets and to the formation of single/binary stars either by contraction or by fission of a single fluid mass.

Modern investigations of the subject became intense in the early seventies (e.g., Toomre 1964; Lebovitz 1972, 1974; Ostriker & Peebles 1973; see also Toomre 1977, 1981, Tassoul 1978, and Durisen & Tohline 1985) when the range of possible applications was extended to the gaseous and stellar disks of our Galaxy and other spiral galaxies, star–forming regions in Giant Molecular Clouds, protostellar and protoplanetary disks, stellar clusters, and elliptical/S0 galaxies. Ostriker & Peebles (1973; hereafter referred to as OP) argued that rapidly rotating, self–gravitating, stellar systems are subject to a violent $m = 2$ nonaxisymmetric instability if $T/|W| \gtrsim 0.14$ (where $T/|W|$ is the ratio of the rotational kinetic energy to the absolute value of the gravitational potential energy), a condition different from $T/|W| \geq 0.2738$ established for incompressible Maclaurin spheroids (EFE). Stellar systems violating the OP criterion were subject to this global instability even if they satisfied the Toomre (1964) criterion for local axisymmetric stability.

The critical values for stability of stellar and fluid systems ($T/|W| \approx 0.14$ and $0.27$, respectively) were confirmed approximately for a variety of models with nonuniform densities and varying angular momentum distributions (e.g., OP; Ostriker & Bodenheimer 1973; Tohline, Durisen, & McCollough 1985; Williams & Tohline 1987, 1988; Miller *et al.* 1989) and were challenged by important discrepancies discovered in some stellar models (Zang 1976; Miller 1978; Miller & Smith 1980) as well as in some fluid models (Tohline & Hachisu 1990; Woodward, Tohline, & Hachisu 1994). After a period of confusion between the point of dynamical instability in stellar systems and the point of secular instability in fluid systems (that occurs at $T/|W| = 0.1375$ in Maclaurin spheroids) it was realized that the instabilities that set in at $T/|W| \approx 0.14$ and $0.27$ in stellar and fluid systems, respectively, are both dynamical (for details see Lebovitz 1972; Bardeen 1975; Bertin & Radicati 1976; Tassoul 1978; Vandervoort & Welty 1982; Durisen & Tohline 1985). The difference between critical values was taken to imply that stellar and fluid systems have important dynamical differences. Anisotropic "pressure" support in stellar systems was particularly to blame (e.g. Hunter 1979).

These conclusions are supported by recent work (Christodoulou, Kazanas, Shlosman, & Tohline 1994; hereafter referred to as CKST). In particular, it was shown that the anisotropic stress–tensor gradients in stellar systems destroy circulation on a dynamical time scale driving a dynamical instability beyond the bifurcation point of elliptical/ellipsoidal Jacobi stellar systems. In fluid systems, the same instability is necessarily secular because it is driven by viscous dissipation which generally destroys circulation over long time scales. On the other hand, the dynamical instability on the fluid Maclaurin sequence beyond the bifurcation point $T/|W| = 0.2738$ of the so–called $x = +1$ self–adjoint Riemann sequence of S–type ellipsoids occurs because circulation is automatically conserved along with mass and angular momentum between objects belonging to the two sequences. These considerations point to two separate kinds of dynamical instability that differ from each other in the way circulation behaves during evolution.

Some investigators, especially those who found counter–examples to the OP criterion, argued that $T/|W|$ does not have a wide physical meaning and, therefore, it cannot serve as a rigorous stability indicator. The fact that it worked well for some models was understood only intuitively. The ratio $T/|W|$ increases with increasing rotation, therefore at some high degree of rotational support instability would set in. Lacking, however, physical justification the $T/|W|$ criteria were termed "semi–empirical." The results presented in CKST provide a physical description of the secular and dynamical instabilities in axisymmetric fluid and stellar systems and suggest that the $T/|W|$ criteria, which are based on an energy ratio, are not sufficiently general to predict all cases of $m = 2$ nonaxisymmetric instability even among axisymmetric systems (see also Efstathiou, Lake, & Negroponte 1982).

If the ratio $T/|W|$ is not the appropriate stability indicator, which parameter may be and how do its critical values compare between fluid and stellar systems? In this paper, we investigate these questions. Motivated by the physical picture presented in CKST and along a line suggested by Vandervoort (1982, 1983), we seek a stability indicator that is based on the angular momentum content rather than on the energy content of a fluid or stellar system. Our plan is to use previously published results about the equilibrium and stability of uniformly rotating, self–gravitating, Maclaurin systems in order to establish a new criterion for stability against $m = 2$ modes and to test the new stability indicator using published axisymmetric models with *different geometries, differential rotation*, and *nonuniform density distributions*. A generalization of the criterion in the case of nonaxisymmetric systems will be presented in a forthcoming paper (Christodoulou, Shlosman, & Tohline 1994).

In §2, we discuss the equilibrium structure, global properties, and stability characteristics of gaseous and stellar disks and spheroids in uniform rotation. Several previous investigations, summarized by Binney & Tremaine (1987), serve as a guide in this presentation but the results are given here in a compact, unified form. In §3, we use these results to formulate a new stability criterion against $m = 2$ nonaxisymmetric modes for uniformly rotating systems. We discuss a physical basis for the new criterion as well as similarities and differences between neutral (viz. stable oscillatory) toroidal modes in stellar and gaseous systems and we argue that the unstable bar modes (OP) of the former and the $m = 2$ Intermediate or I–modes (Goodman & Narayan 1988) of the latter models are basically similar. In this section, we also discuss various stability indicators that have been previously proposed as extensions or replacements for the ratio $T/|W|$ of the OP criterion. In §4, we demonstrate that the new stability criterion also holds for a variety of stellar/gaseous models with different geometries, differential rotation, and nonuniform density distributions. In §5, we summarize our conclusions.

## 2  MACLAURIN SYSTEMS

In this section, we review the equilibrium equations and the characteristic equations for $m = 2$ nonaxisymmetric modes in uniformly rotating, self–gravitating, gaseous and stellar, axisymmetric disks and spheroids. We also introduce global dynamical quantities such as the rotational kinetic energy and the gravitational potential energy as well as the appropriate Jeans frequency due to self–gravity. Some of the following results can also be found in the book "Galactic Dynamics" by Binney & Tremaine (1987).

### 2.1  Gaseous Maclaurin Spheroid

The Maclaurin spheroid is presented in detail in EFE. It is an axisymmetric model of an incompressible fluid with uniform density $\rho$ and uniform rotation of angular velocity $\Omega$. The eccentricity $e$ of the meridional sections is given by the equation

$$e = \left(1 - \frac{a_3^2}{a_1^2}\right)^{1/2}, \tag{2.1}$$

where $a_1$ and $a_3$ are the principal axes of the spheroid. For our purposes, we consider oblate spheroids with $a_3 \leq a_1$. (Prolate spheroidal equilibria with constant density and uniform rotation do not exist; see Florides & Spyrou 1993.) The angular velocity and the eccentricity are related through Maclaurin's formula

$$\frac{\Omega^2}{\pi G \rho} = \frac{2(1-e^2)^{1/2}}{e^3}(3 - 2e^2)\sin^{-1} e - \frac{6}{e^2}(1 - e^2), \tag{2.2}$$

where $G$ is the gravitational constant. The total mass $M$, kinetic energy due to rotation $T$, and gravitational potential energy $W$ are given by the equations

$$M = \frac{4\pi}{3}\rho a_1^3 (1 - e^2)^{1/2}, \tag{2.3}$$

$$T = \frac{1}{5} M \Omega^2 a_1^2, \tag{2.4}$$

and

$$W = -\frac{3}{5}\frac{GM^2}{a_1}\frac{\sin^{-1} e}{e}, \tag{2.5}$$

respectively. Combining equations (2.2)–(2.5), we find that the ratio $t \equiv T/|W|$ is a function of the eccentricity only (e.g., Bodenheimer & Ostriker 1973), i.e.

$$t = \frac{3}{2e^2}\left[1 - \frac{2}{3}e^2 - \frac{e}{\sin^{-1} e}(1 - e^2)^{1/2}\right]. \tag{2.6}$$

Dynamical instability sets in at $e = 0.95289$ (EFE) which corresponds to $t = 0.27383$.

The gravitational potential at any interior point is harmonic, i.e.

$$\Phi(R, Z) = \pi G \rho \left[ A_1(R^2 - 2a_1^2) + A_3(Z^2 - a_3^2) \right], \tag{2.7}$$

where $R$ and $Z$ denote cylindrical coordinates and

$$A_1 = \frac{(1-e^2)^{1/2}}{e^3}\sin^{-1}e - \frac{1-e^2}{e^2}, \tag{2.8}$$

and

$$A_3 = 2(1 - A_1) = \frac{2}{e^2} - \frac{2(1-e^2)^{1/2}}{e^3}\sin^{-1}e. \tag{2.9}$$

The Jeans frequencies due to self–gravity in $R$ and $Z$ are now introduced naturally through equation (2.7). For the nonaxisymmetric modes of interest, we shall find useful the Jeans frequency $\Omega_J$ in the radial direction

$$\Omega_J^2 = 2\pi G\rho A_1. \tag{2.10}$$

As is demonstrated in §2.2 below, this is the appropriate gravitational frequency to use for both disks and spheroids — not just the dimensionally correct term $\pi G\rho$. Combining equations (2.3)–(2.5) and (2.10), we find another useful expression for $t$, i.e.

$$t = \frac{1}{2}f\Big(\frac{\Omega}{\Omega_J}\Big)^2, \tag{2.11}$$

where

$$f \equiv \frac{A_1}{\sqrt{1-e^2}}\frac{e}{\sin^{-1}e} = \frac{1}{e^2}\Big[1 - \frac{e}{\sin^{-1}e}\sqrt{1-e^2}\Big]. \tag{2.12}$$

From equations (2.6) and (2.12), we find that $t = \frac{3}{2}f - 1$. Notice that $f$ varies only between 2/3 ($e=0$, $t=0$, sphere) and 1 ($e=1$, $t=1/2$, disk). Using now $e=0.95289$ and $t=0.27383$, we find that $f=0.84922$ and that dynamical instability sets in at $\Omega/\Omega_J = 0.80306$.

Linear stability analysis of the incompressible Maclaurin spheroid is also presented in EFE. The characteristic equation for the toroidal modes with frequency $\omega$ can be written in dimensionless form as

$$\sigma^2 \mp 2\chi\sigma - (b_{11} - 2\chi^2) = 0, \tag{2.13}$$

from which we find that

$$\sigma = \pm\chi \pm (b_{11} - \chi^2)^{1/2}, \tag{2.14}$$

where

$$\sigma \equiv \frac{\omega}{\Omega_J}, \tag{2.15}$$

$$\chi \equiv \frac{\Omega}{\Omega_J}, \tag{2.16}$$

and $\Omega_J$ is given by equation (2.10). In equations (2.13) and (2.14), $b_{11}$ is a function of the eccentricity only, i.e.

$$b_{11} = \frac{2B_{11}}{A_1}, \tag{2.17}$$

where

$$B_{11} = \frac{\sqrt{1-e^2}}{4e^4}\Big[(3-2e^2)\sqrt{1-e^2} - (3-4e^2)\frac{\sin^{-1}e}{e}\Big], \tag{2.18}$$

and $A_1$ is given by equation (2.8).

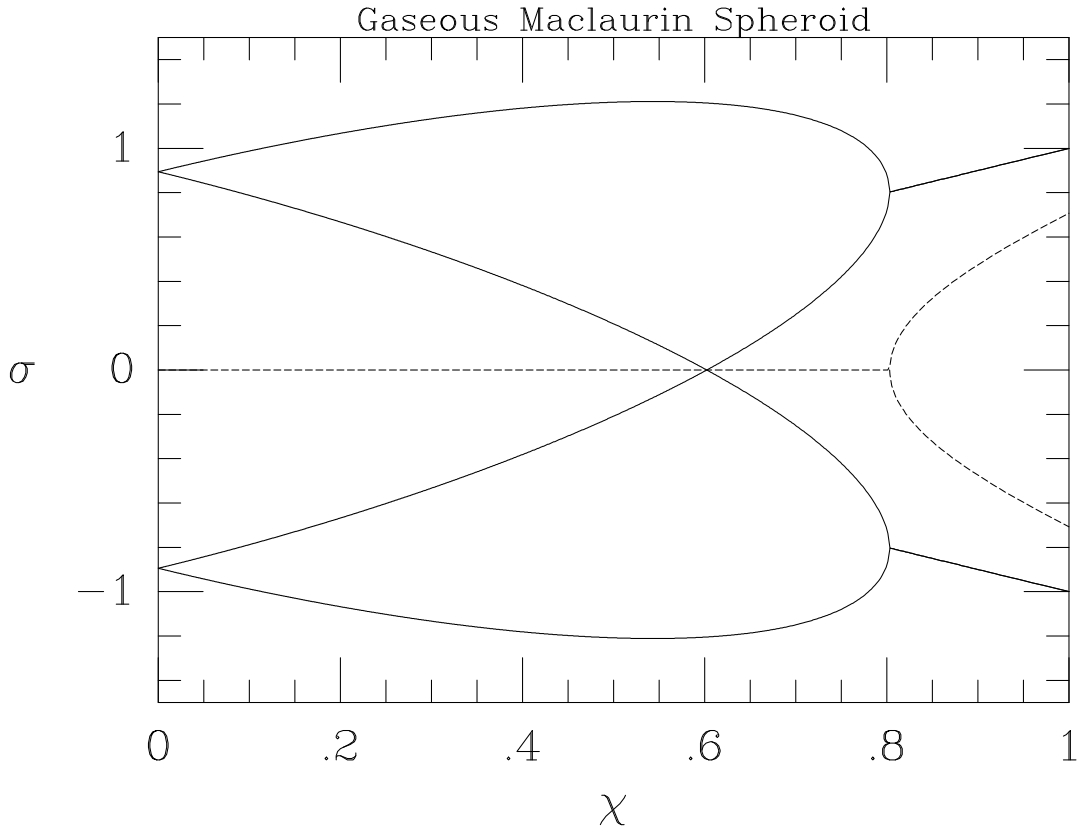

FIGURE 1. *Branches of toroidal modes in the gaseous Maclaurin spheroid (see also EFE). The real and imaginary parts of the eigenfrequency $\sigma$ are plotted as solid and dashed lines, respectively.*

The four modes described by equation (2.14) are plotted in Figure 1 as functions of $\chi$. The $\sigma \geq 0$ branches are also plotted in EFE against eccentricity, an equivalent parameter to $\chi$. (Note, however, that $\omega^2$ in EFE is given in units of $\pi G\rho$.) The growth rates of the unstable modes are also plotted in Figure 1 as dashed lines. Two neutral branches cross the $\sigma = 0$ line at $\chi = 0.60225$ ($e=0.81267$), the bifurcation point of the Jacobi ellipsoids. As mentioned above, dynamical instability appears at $\chi = 0.80306$ ($e=0.95289$). Each unstable branch is a merger of two neutral branches. The two merging neutral modes carry equal but opposite in sign amounts of perturbed angular momentum (see Christodoulou & Narayan 1992 and references therein). Looking at the upper half of the diagram, the modes lying in the upper (high $\sigma$) branch are called "fast" while those lying in the lower (low $\sigma$) branch are called "slow." Right before merging, both branches have positive frequencies, i.e. their modes are prograde relative to the direction of rotation of the spheroid. What is most interesting in Figure 1 (and not so obvious in the EFE plot) is that the "slow" neutral branch is neither slow nor prograde for all values of $\chi < 0.80306$. In fact, this branch is made of fast retrograde modes in the limit $\chi \to 0$. The branch of retrograde modes crosses to positive $\sigma$ at the Jacobi bifurcation point and becomes prograde for $\chi > 0.60225$. Such behavior of the "slow" branch is not seen in very slender rotating annuli with no pressure support (Christodoulou & Narayan 1992) or in stellar systems (see below). It is only similar to the behavior of the "slow" modes in gaseous Maclaurin disks described in §2.2 below. We shall discuss these neutral branches in §3.1.

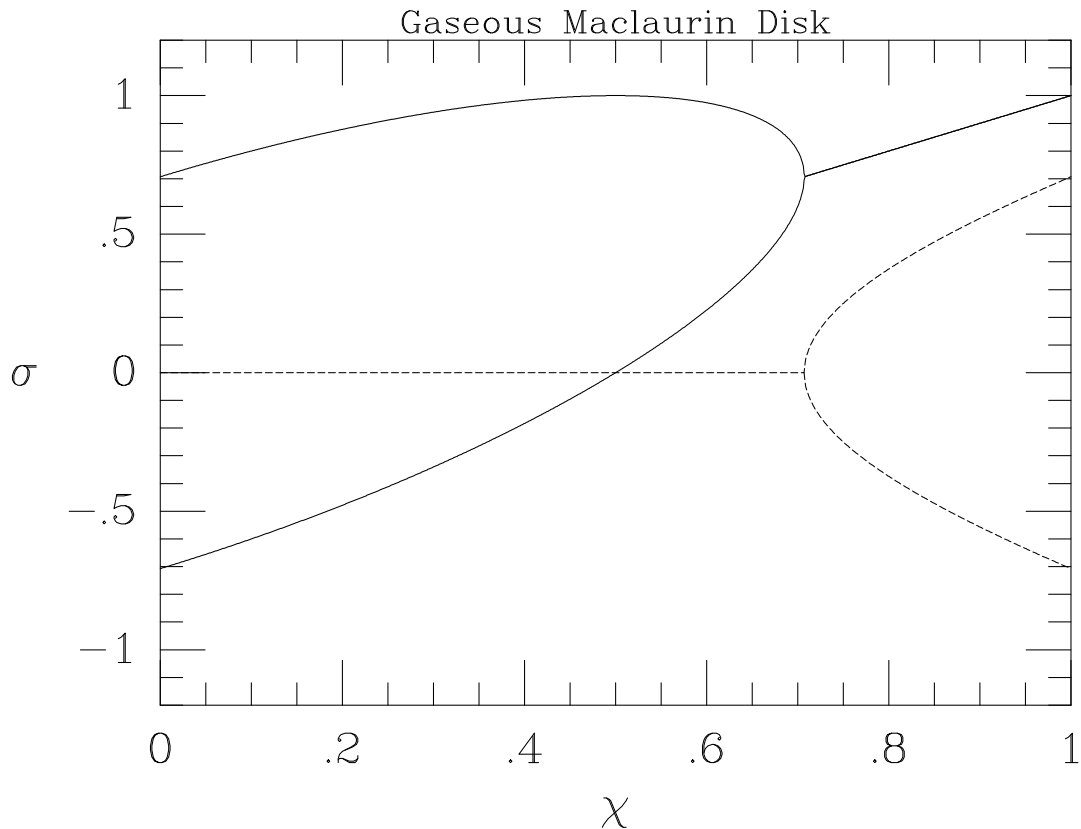

FIGURE 2. *Branches of toroidal modes in the gaseous Maclaurin disk (see also Hunter 1979). The real and imaginary parts of the eigenfrequency $\sigma$ are plotted as solid and dashed lines, respectively.*

### 2.2 Gaseous Maclaurin Disk

A brief description of this model can be found in Hunter (1979) along with a comparison between modes in gaseous and stellar Maclaurin disks. The Maclaurin disk is axisymmetric, self–gravitating, rotates uniformly with angular velocity $\Omega$, and has surface density of the form

$$\Sigma(R) = \Sigma_o \Big(1 - \frac{R^2}{R_o^2}\Big)^{1/2}, \qquad (2.19)$$

where $R$ is the radial cylindrical coordinate, $R_o$ is the radius of the disk, and $\Sigma_o$ is the central surface density. The gravitational potential at any point in the disk is again harmonic, i.e.

$$\Phi(R) = \frac{1}{2}\Omega_J^2\Big(R^2 - 2R_o^2\Big), \qquad (2.20)$$

where $\Omega_J$ is again defined by equation (2.10) in the limit $e=1$ (disk). For $e=1$, equation (2.10) written in terms of the total mass $M$ reduces to

$$\Omega_J^2 = \frac{3\pi GM}{4R_o^3}, \tag{2.21}$$

exactly as was defined by Kalnajs (1972) and Hunter (1979). We see therefore that the definition (2.10) of the Jeans frequency is quite general and applies uniformly to all Maclaurin systems. (For systems with different density distributions, it is not obvious what the relevant Jeans frequency is. We shall return to this point in §4 below.)

The total mass $M$, kinetic energy due to rotation $T$, and gravitational potential energy $W$ are given by the equations

$$M = \frac{2\pi}{3}\Sigma_o R_o^2, \tag{2.22}$$

$$T = \frac{1}{5}M\Omega^2 R_o^2, \tag{2.23}$$

and

$$W = -\frac{2}{5}M\Omega_J^2 R_o^2, \tag{2.24}$$

respectively. Combining equations (2.23) and (2.24), we find that the ratio $t \equiv T/|W|$ is given by the equation

$$t = \frac{1}{2}\left(\frac{\Omega}{\Omega_J}\right)^2. \tag{2.25}$$

in agreement with equation (2.11) in the disk limit $e=1$.

The characteristic equation for $m = 2$ toroidal modes with frequency $\omega$ (Hunter 1979) can be written in dimensionless form in the inertial frame as

$$\sigma^2 - 2\chi\sigma + 2\chi^2 - \frac{1}{2} = 0, \tag{2.26}$$

where $\sigma$ and $\chi$ are defined again by equations (2.15) and (2.16), respectively. From equation (2.26), we find that

$$\sigma = \chi \pm \left(\frac{1}{2} - \chi^2\right)^{1/2}. \tag{2.27}$$

These two branches of the characteristic equation are plotted in Figure 2 as functions of $\chi$. Hunter (1979) also plots the same modes but the frequency $\omega_H$ is evaluated in a rotating frame such that $\omega_H = \omega - 2\Omega$. Figure 2 is similar to Figure 1. (The second pair of branches that appears in Figure 1 is not plotted here for clarity.) Notice again that the "slow" branch is composed of fast retrograde modes at low rotation frequencies. This branch crosses the $\sigma = 0$ line at $\chi = 1/2$, the bifurcation point of elliptical Jacobi disks for which $t = 1/8$ [equation (2.25); cf. Weinberg & Tremaine (1983)].

Dynamical instability appears at $\chi = 1/\sqrt{2} = 0.70711$ as a merger of two prograde neutral modes. Using this value in equation (2.25) we find that $t = 1/4$. The maximum growth rate in both Figures 1 and 2 is obtained for $\chi = 1$ and is $\sigma_I = 1/\sqrt{2}$. The reason for this equivalence is that if $\chi = 1$, then $e=1$ and the Maclaurin spheroid reduces exactly to the Maclaurin disk. In this case, $b_{11} = 1/2$ and equation (2.13) reduces to equation (2.26).

### 2.3 Stellar Maclaurin Disk ($\Omega$–model)

The $\Omega$–model (the stellar analog of the gaseous Maclaurin disk) was studied for stability by Kalnajs (1972). A subsequent study by Kalnajs & Athanassoula (1974) focused on the $m = 2$ bar modes. Hunter (1979) studied this model in the context of stellar hydrodynamics and compared the bar modes to the toroidal modes of the gaseous disk. The $\Omega$–model is a uniformly rotating, self–gravitating, axisymmetric, stellar disk whose surface density, potential, Jeans frequency, and global properties are given by equations (2.19)–(2.25) above. The choaracteristic equation for $m = 2$ modes is, however, different from equation (2.26) because of anisotropic "pressure" support in the perturbations of the stellar model (Hunter 1979). Following Kalnajs & Athanassoula (1974), we write in dimensionless form

$$\sigma^3 - \frac{5}{2}\sigma + 3\chi = 0, \tag{2.28}$$

where $\sigma$ and $\chi$ are defined again by equations (2.15) and (2.16), respectively, and $\chi$ represents now the dimensionless mean rotation frequency. The three branches of the characteristic equation are plotted in Figure 3 as functions of $\chi$.

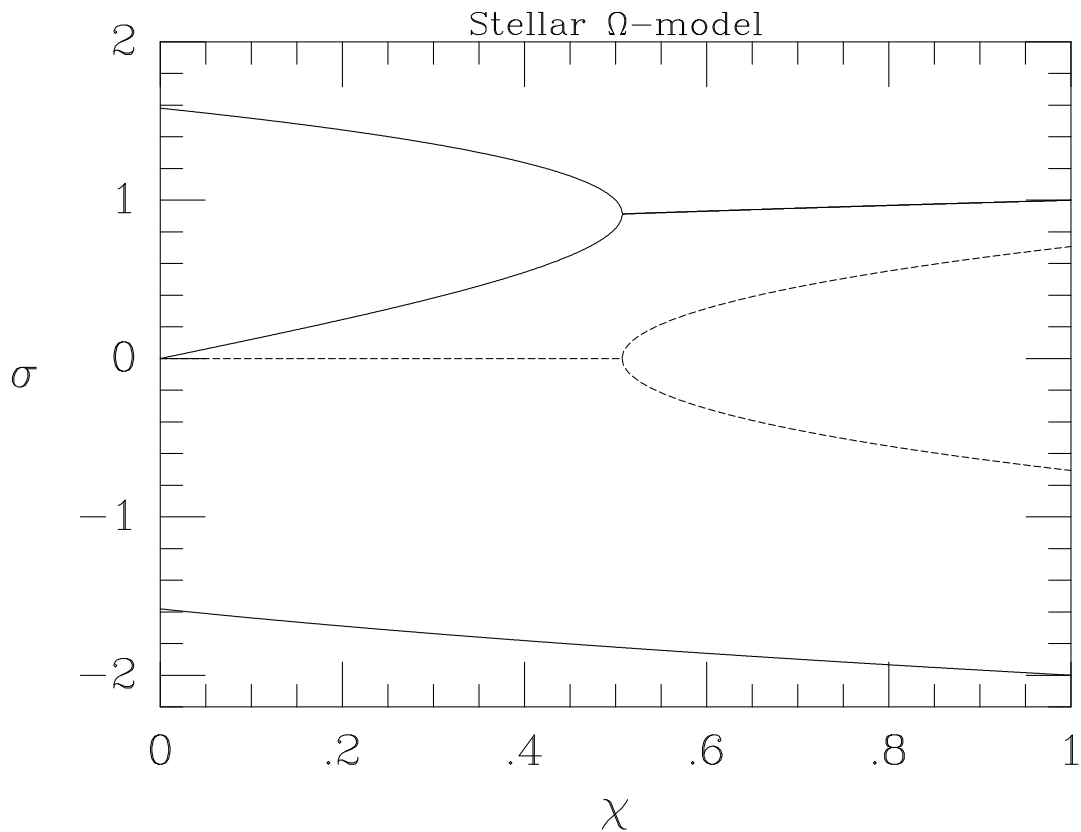

FIGURE 3. Branches of $m = 2$ bar modes in the stellar Maclaurin disk (see also Kalnajs & Athanassoula 1974 and Hunter 1979). The real and imaginary parts of the eigenfrequency $\sigma$ are plotted as solid and dashed lines, respectively.

This plot was also presented by Kalnajs & Athanassoula (1974) who showed that the "fast" and "slow" prograde branches carry opposite in sign angular momentum perturbations (positive and negative, respectively). Unlike in the previous two gaseous models, the slow branch is always prograde in the $\Omega$–model. Two modes merge at $\chi = 5(5/6)^{1/2}/9 = 0.50715$ and produce the unstable bar mode. Using this value in equation (2.25), we find that $t = 0.12860$. The maximum growth rate is obtained for $\chi = 1$ and is again $\sigma_I = 1/\sqrt{2}$. The branch of retrograde modes seen at $\sigma < 0$ is always neutral and is not involved with the prograde bar modes.

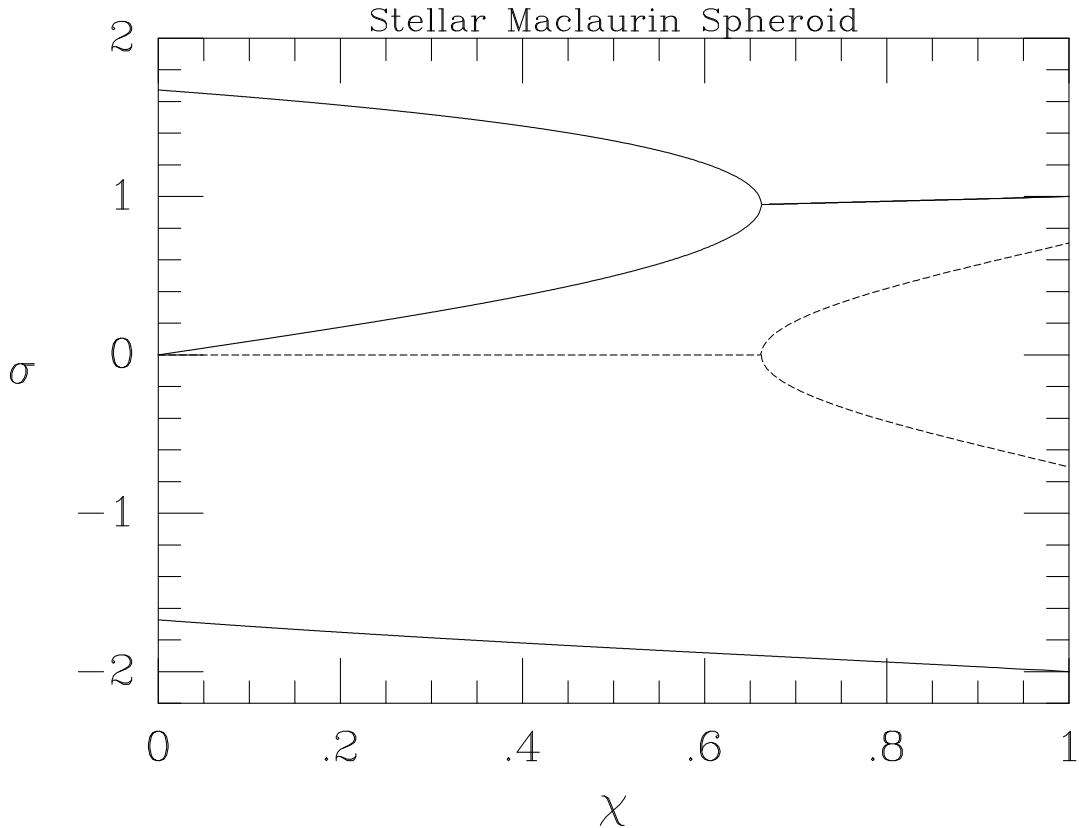

FIGURE 4. *Branches of toroidal modes in the stellar Maclaurin spheroid (see also Vandervoort 1991). The real and imaginary parts of the eigenfrequency $\sigma$ are plotted as solid and dashed lines, respectively.*

### 2.4  Stellar Maclaurin Spheroid

A family of uniformly rotating, self–gravitating, homogeneous, stellar spheroids was constructed and studied for stability by Vandervoort (1991). The mean angular velocity $\Omega$ is a free parameter in these models. Adopting equation (2.2) to relate $\Omega$ to the eccentricity $e$, we consider a special class of stellar spheroids analogous to the incompressible Maclaurin spheroids. The characteristic equation for toroidal modes with frequency $\omega$ in the inertial frame is a cubic equation; written in dimensionless form it reads

$$\sigma^3 - (2 + b_{11})\sigma \pm 2(2 - b_{11})\chi = 0, \qquad (2.29)$$

where $\sigma$, $\chi$, and $b_{11}$ are defined by equations (2.15), (2.16), and (2.17), respectively. We have obtained equation (2.29) from equation (36) of Vandervoort (1991) using the identity $B_{11} = A_1 - a_1^2 A_{11}$ (EFE). In the disk limit $e=1$, $b_{11} = 1/2$ and equation (2.29) reduces to the characteristic equation (2.28) for the stellar Maclaurin disk.

The three branches corresponding to the plus sign of equation (2.29) are plotted in Figure 4 as functions of $\chi$. Figure 4 is similar to Figure 3. Dynamical instability appears at $\chi = 0.66212$ ($e=0.86362$) as a merger between two prograde neutral modes. The maximum growth rate is obtained for $\chi = 1$ and is again $\sigma_I = 1/\sqrt{2}$. The slow branch that is involved with the instability is always prograde. The branch of retrograde modes seen at $\sigma < 0$ is always neutral and is not involved with the prograde bar modes.

Equations (2.1)–(2.12) are also valid for the stellar Maclaurin spheroid. Using $e=0.86362$ in equations (2.6) and (2.12), we find that $f=0.78076$ and that dynamical instability sets in at $t = 0.17114$.

## 3  The Stability Parameter $\alpha$

### 3.1  Neutral and Unstable Modes

For comparison purposes, we plot as functions of $\chi$ the real parts of the two branches that produce the unstable modes in spheroids (Figure 5) and disks (Figure 6). The critical values at which dynamical instability sets in are also summarized in Table 1 along with the fractional variations in $\chi = \Omega/\Omega_J$ and $t = T/|W|$ between disks and spheroids. A comparison leads to the following differences: (a) Instability sets in at lower values of $\chi$ (lower rotation) in stellar models than in gaseous models. (b) Even for stellar or gaseous systems separately, instability sets in at different values of $\chi$. Thus, the onset of instability depends on an additional parameter, the geometric shape of the system, and the ratio $\Omega/\Omega_J$ alone cannot serve as a solid stability indicator for Maclaurin systems. The same argument applies to the ratio $T/|W|$. (c) The lower neutral branches of the gaseous systems begin as "fast" and retrograde at low rotations but eventually cross over to positive $\sigma$ at the corresponding Jacobi bifurcation points and continue to increase substantially in frequency with increasing rotation before merging with the corresponding "fast" prograde branches. No such behavior occurs in stellar systems.

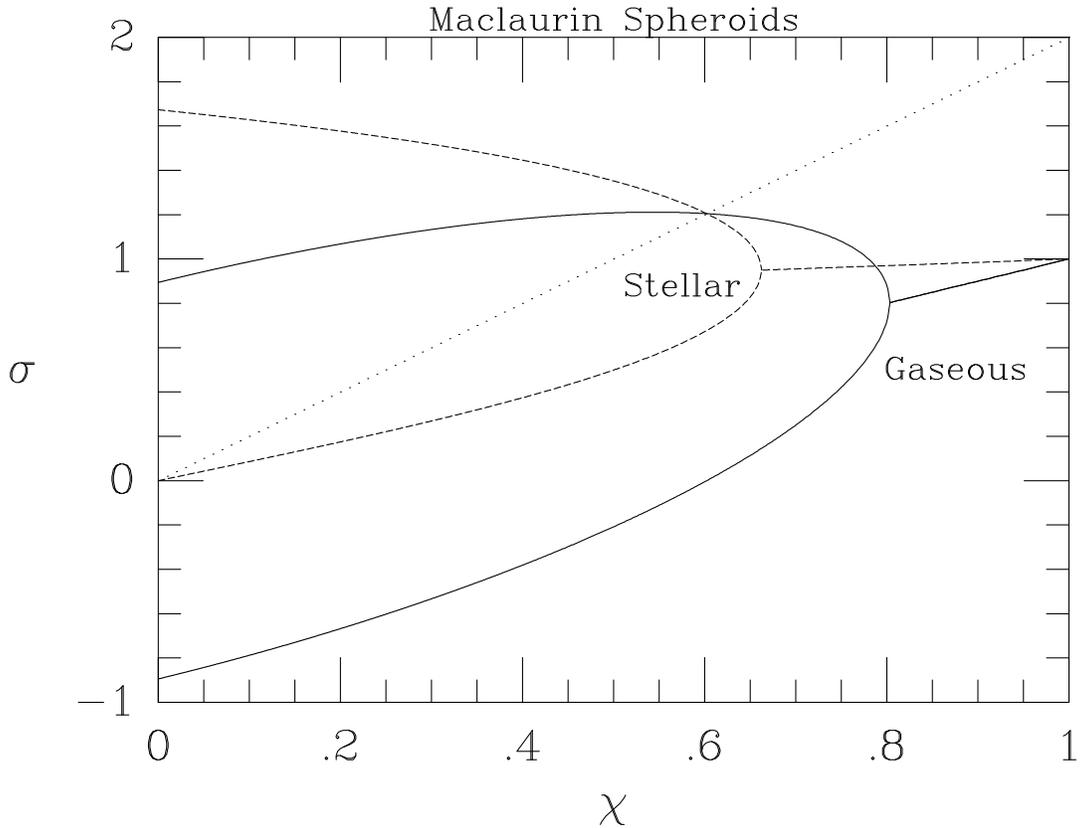

FIGURE 5. *Comparison between $m = 2$ modes in stellar and gaseous Maclaurin spheroids. Solid lines denote branches in the gaseous spheroid while dashed lines denote branches in the stellar spheroid. Only the real part of the eigenfrequency $\sigma$ is plotted. The two systems have a common fast neutral mode at $\chi = 0.60225$ (the bifurcation point of gaseous Jacobi ellipsoids where the slow neutral mode has $\sigma = 0$). This common mode corotates with the spheroid. Corotation corresponds to $\sigma = 2\chi$ and is denoted by a dotted line.*

This last difference, however, has no physical significance since the definitions of the mean angular velocity $\Omega$ are not the same in the two types of systems. Specifically, in stellar systems, figure rotation is described by a rotation of the coordinate frame with mean frequency $\Omega \leq 0$ while the circulation due to azimuthal motions is not included to the mean frequency (e.g. Freeman 1966a,b,c; Hunter 1974). In gaseous Maclaurin systems, $\Omega$ denotes the mean rotation frequency of the equilibrium figures in the inertial frame (EFE). Since the rotation is uniform, there are no vortical motions in a frame rotating with frequency $\Omega$ and the circulation in a gaseous Maclaurin system is zero.

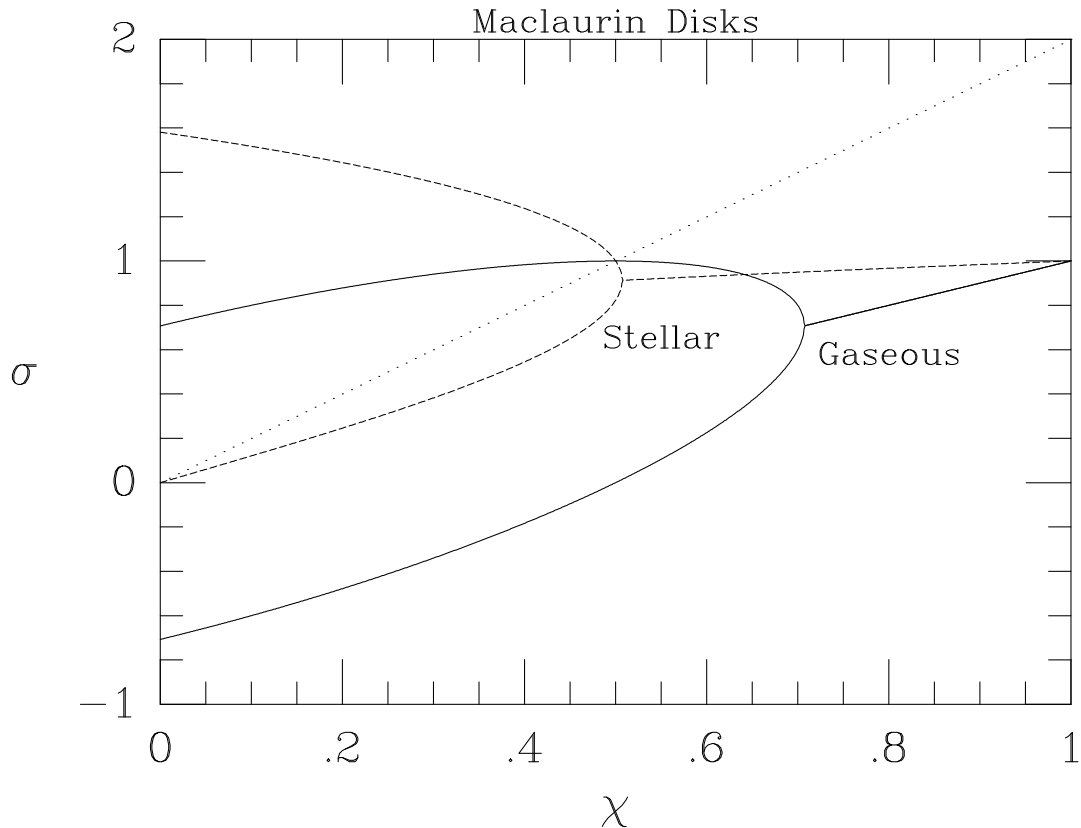

FIGURE 6. *Comparison between $m = 2$ modes in stellar and gaseous Maclaurin disks (see also Hunter 1979). Solid lines denote branches in the gaseous disk while dashed lines denote branches in the stellar disk. Only the real part of the eigenfrequency $\sigma$ is plotted. The two systems have a common fast neutral mode at $\chi = 1/2$ (the bifurcation point of gaseous elliptical Jacobi disks where the slow neutral mode has $\sigma = 0$). This common mode corotates with the disk. Corotation corresponds to $\sigma = 2\chi$ and is denoted by a dotted line.*

We can compensate for this difference and thus obtain a meaningful comparison between neutral modes in gaseous and stellar systems as follows. We assume that neutral modes in gaseous Maclaurin systems are viewed from a rotating frame of angular velocity $\Omega_{fr} \leq 0$ such that $\sigma_{fr} \equiv \Omega_{fr}/\Omega_J$ and the dimensionless mode frequencies $\sigma_R \equiv \omega'/\Omega_J$ in this frame are given by $\sigma_R = \sigma - \sigma_{fr}$. This is equivalent to assuming the existence of a mean circulation in gaseous systems superimposed on the mean rotation which is now expressed through the frame's rotation. We adopt $\sigma_{fr} = \chi - 1/\sqrt{2} \leq 0$ for disks and $\sigma_{fr} = \chi - \sqrt{b_{11}} \leq 0$ for spheroids so that $\sigma_{fr} = 0$ at the points of onset of instability. Then, the dimensionless mode frequencies $\sigma_R$ are

$$\sigma_R = \sqrt{b_{11}} \pm \sqrt{b_{11} - \chi^2} \qquad \text{(gaseous spheroids)}, \tag{3.1}$$

and

$$\sigma_R = \frac{1}{\sqrt{2}} \pm \sqrt{\frac{1}{2} - \chi^2} \qquad \text{(gaseous disks)}. \tag{3.2}$$

These equations should be contrasted to equations (2.14) and (2.27), respectively.

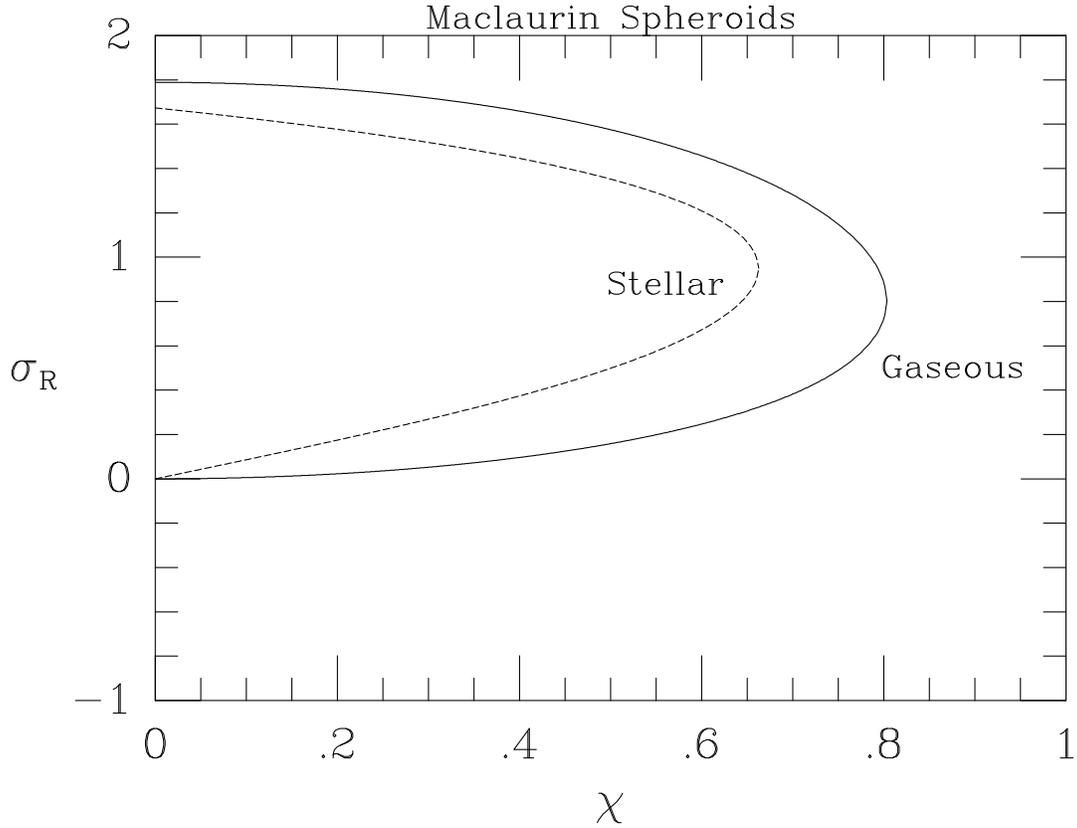

FIGURE 7. Comparison between neutral $m = 2$ modes in stellar and gaseous Maclaurin spheroids. The real eigenfrequencies $\sigma_R$ in the rotating frame (solid line) are plotted versus $\chi$ for gaseous spheroids using equation (3.1). The neutral modes in stellar spheroids (dashed line) are unchanged from Figure 5, i.e. $\sigma_R = \sigma$. The frame transformation produces a zero–frequency mode ($\sigma_R = 0$) at $\chi = 0$ and leaves unchanged ($\sigma_R = \sigma$) the marginally stable modes at $\chi = 0.80306$ in gaseous spheroids.

Figure 7 shows the result from the frame transformation in gaseous Maclaurin spheroids. The neutral branches of stellar spheroids are the same as in Figure 5. The neutral branches of gaseous spheroids are transformed so that the point of marginal stability at $\chi = 0.80306$ remains unchanged. In addition, the slow branch starts from $\sigma_R = 0$ at $\chi = 0$ and is now entirely prograde. The similarity between neutral branches depicted in Figure 7 is also exhibited by the neutral branches in Maclaurin disks when the modes of gaseous disks are given by equation (3.2) in a rotating frame.

Three interesting clues about the nature of the neutral modes can be deduced from Figure 7: (1) The fast branches are similar in shape but exist at somewhat different frequencies even for $\chi = 0$. This indicates that these modes are only quantitatively different in the two types of systems after the frame transformation. (2) The slow stellar branch increases in frequency more steeply than the slow gaseous branch (see below). (3) The most important difference is in the location of the onset of dynamical instability: gaseous systems are more stable than stellar systems, i.e. instability sets in first in stellar systems with increasing $\chi$ (see also Binney & Tremaine 1987; for more details see CKST).

However, the resulting unstable modes have similar properties (see Figures 5 and 6). Their growth rates are quite similar (§2) and their pattern speeds are smaller than $\Omega$. The pattern speeds are exactly equal to $\Omega/2$ in gaseous models and approximately equal to $\Omega/2$ in stellar models. Previous studies of uniformly *and* differentially rotating models also indicate that, irrespective of the type of the system, unstable (stellar) bar modes and $m = 2$ (fluid) I–modes have the same linear and nonlinear characteristics. (For stellar modes see e.g. Kalnajs 1971, 1972, 1978; Lynden–Bell & Kalnajs 1972; Tremaine 1976; Zang & Hohl 1978; Hunter 1979; Toomre 1981; Sellwood 1983; Athanassoula 1984; Sellwood & Athanassoula 1986. For fluid modes see e.g. EFE; Lebovitz 1972; Tohline, Durisen, & McCollough 1985; Williams & Tohline 1987, 1988; Goodman & Narayan 1988; Tohline & Hachisu 1990; Christodoulou & Narayan 1992; Christodoulou 1993; Woodward *et al.* 1994.) For example, in the linear regime, unstable systems are deformed to elliptical/ellipsoidal shapes. When perturbations grow to a nonlinear amplitude, all systems appear as elongated bars with trailing spiral arms at the bar edges. The unstable modes do not appear to be dependent on or driven by the corotation resonance (Goodman & Narayan 1988; Christodoulou 1993). In all cases, the pattern speed deduced from numerical simulations is less than the characteristic orbital frequency $\Omega$ and rather close to the value $\Omega/2$. As a result, the linear size of the bar continues to grow with time (cf. Sellwood 1981; Efstathiou, Lake, & Negroponte 1982; Christodoulou & Narayan 1992). The spiral arms eventually dissipate away after winding up in purely gaseous and stellar systems (e.g., Hohl 1971; Athanassoula & Sellwood 1986; Williams & Tohline 1988).

Returning now to Figure 7 and to difference (2) above, the neutral modes appear to be driven primarily by self–gravity and rotation (cf. Goodman & Narayan 1988; Christodoulou & Narayan 1992). In the limit $\chi \to 0$, equations (2.28) and (2.29) imply that $\omega_F \sim \Omega_J$ and $\omega_S \sim \Omega$ in stellar systems. (The indices $F$ and $S$ denote the fast and slow modes, respectively.) In the same limit, equations (3.1) and (3.2) along with the definition $\sigma_R \equiv \omega'/\Omega_J$ imply that $\omega'_F \sim \Omega_J$ and $\omega'_S \sim \Omega^2/\Omega_J$ in gaseous systems. Thus, the fast prograde modes appear to be supported by self–gravity at low rotations. The slow prograde modes appear to be supported primarily by rotation; they have zero frequency at $\Omega = 0$ and their frequencies increase as $\Omega$ increases. (In the same context, the fast, always neutral, retrograde modes in stellar systems remain basically unaffected by rotation as $\Omega$ increases.) We do not understand the origin of the difference at low rotations between the transformed slow modes in gaseous systems ($\omega'_S \sim \Omega^2/\Omega_J$) and the analogous modes in stellar systems ($\omega_S \sim \Omega$). This difference is ultimately responsible for the onset of instability at lower $\chi$–values in stellar systems. The final outcome is, however, the appearance of an unstable branch at high rotations with the same fundamental characteristics in both types of systems.

### 3.2 Marginal Stability

We now return to the results listed in Table 1. Although there is no unique critical value for stability applicable to both gaseous and stellar Maclaurin systems, one can still argue that such critical values should exist separately for each type of system (cf. CKST). The variation seen in Table 1 is sufficient to exclude the parameters $t$ and $\chi$ in gaseous systems as well as in stellar systems. However, the ratio

$$\alpha \equiv \frac{t}{\chi} = \frac{T/|W|}{\Omega/\Omega_J}, \quad (0 \le \alpha \le 1/2), \tag{3.3}$$

can play the role of the stability indicator because the individual variations in the critical values of $t$ and $\chi$ are nicely compensated. We list in Table 2 the corresponding critical values of the parameter $\alpha$ obtained from the values of Table 1. The critical values are $\alpha \approx 0.34$ for stability of gaseous systems and $\alpha \approx 0.25$ for stellar systems while the variation between disks and spheroids has decreased to less than 4% for gaseous systems and to less than 2% for stellar systems. For Maclaurin systems, we can use equations (2.11), (2.16), and (3.3) to write

$$\alpha = (\frac{1}{2}ft)^{1/2}. \tag{3.4}$$

We note that $f = \frac{2}{3}(1+t)$ and this equation can be written in the alternative form $\alpha = [t(1+t)/3]^{1/2}$.

Although equations (3.3) and (3.4) were established in the simple case of uniformly rotating systems, the underlying physics should by quite general (see §3.3 below). In what follows, we test the applicability of the parameter $\alpha$ as a stability indicator in differentially rotating, centrally condensed, axisymmetric systems. For such complicated systems, it is not clear how appropriately weighted values of $\Omega_J$ and $\Omega$ should be obtained and equation (3.3) cannot be easily applied (but see §4.3 and §4.4a). On the other hand, equation (3.4) is simpler to use for applications and provides an accurate stability indicator for various differentially rotating disk and spheroidal models. We, therefore, shall use equation (3.4) when discussing stability of nonaxisymmetric systems as well (Christodoulou, Shlosman, & Tohline 1994).

### 3.3 Physical Significance

For uniformly rotating systems, parameter $\alpha$ can be interpreted physically as a ratio between specific angular momenta. We demonstrate this most simply for Maclaurin systems for which equation (3.3) can be written as

$$\alpha = \frac{5f}{4} \frac{L/M}{\Omega_J a_1^2}, \tag{3.5}$$

where $L \equiv 2M\Omega a_1^2/5$ is the total angular momentum, $a_1$ represents the equatorial radius, and $\Omega_J, f$ are given by equations (2.10), (2.12), respectively. The term $\Omega_J a_1^2$ is the maximum angular momentum of a circular orbit in the equatorial plane. Since $5f/4 \approx 1$ for spheroids (Table 1) and $f = 1$ for disks, the critical values listed in Table 2 are simply a measure of how much the specific angular momentum $L/M$ can increase relative to $\Omega_J a_1^2$ in a stable Maclaurin system.

Equation (3.4) also has topological significance. For the uniformly rotating Maclaurin systems, the new stability indicator is a simple combination of the commonly used ratio $T/|W|$ and the function $f$ that depends on both the topology and the geometry of a system. Introducing the topology of a system explicitly into the stability indicator is not surprising. For example, Tohline & Hachisu (1990) find that rotating, self–gravitating gaseous tori are subject to an unstable $m = 2$ I–mode if $T/|W| \gtrsim 0.16-0.17$. This critical value is significantly lower than the familiar value of 0.27 in spheroids and it would seem that the difference has its origin in the change of topology. Along similar lines, Vandervoort (1983) finds a substantial variation in the critical values of the ratio $T/|W|$ in stellar spheroidal families whose structure ranges from a sphere to a disk (see §3.4a below).

Vandervoort (1982, 1983) has suggested that the angular momentum $L$ rather than the kinetic energy $T$ should be the quantity that generally decides stability in rotating self-gravitating systems. In the spirit of this suggestion, $T$ should be replaced with $L^2/2I$, where $I$ is the moment of inertia about the symmetry axis of a system. Following this idea and using $\Omega = L/I$ in equation (3.3), we find that

$$\alpha = T_J/|W|, \tag{3.6}$$

where

$$T_J \equiv \frac{1}{2} L \Omega_J. \tag{3.7}$$

Again, this expression is strictly valid for uniformly rotating systems but the suggested physical meaning — to shift the focus from $T$ to the product $L\Omega_J$ — should have wider implications (as we have argued also for its counterpart — equation [3.4]) in view of the importance of the angular momentum and self–gravity in determining the onset of instability (§3.1; CKST).

### 3.4 Comparison with Previous Stability Criteria in Stellar Systems

Here we compare the new stability criterion with criteria derived previously from detailed studies of particular models. In general, all studied models are stellar and differentially rotating.

(a) *Spheroidal Models*: Families of purely self–gravitating, differentially rotating, stellar spheroids with different eccentricities were considered by Vandervoort (1983). In the limit of uniform (or averaged) rotation, the parameter $t$ for marginal stability varies between 0.1286 ($e$=1 family) and 0.1882 ($e$=0 family). The critical values of the parameter $\alpha$ evaluated from equation (3.4) vary much less than $t$, taking the minimum value $\alpha = 0.2505$ for the $e$=0 family and the maximum value $\alpha = 0.2594$ for the $e$=0.95 family. These critical values of $\alpha$ are in good agreement with the results listed in Table 2.

(b) *Toomre/Plummer Disk/Halo Models*: Athanassoula & Sellwood (1986, 1987) studied numerically the stability of stellar models composed of a Kuzmin–Toomre disk (see e.g. Toomre 1962) embedded in a Plummer (1911) halo. Frank & Shlosman (1989) demonstrated analytically that the marginal stability line produced by the numerical experiments agrees reasonably well with the value $t = 0.14$ predicted by the OP criterion. Using $t=0.14$ and $f=1$ in equation (3.4) we find that $\alpha = 0.26$, a value that does imply marginal stability as well. We conclude that the OP criterion and the parameter $\alpha$ both describe accurately the stability of Toomre/Plummer disk/halo equilibrium systems. Additional composite models are discussed in §4.5 below.

*(c) Finite–Thickness Disk/Halo Models*: A stability criterion for flattened stellar systems of nonzero vertical thickness $h$ and mass $M_D$ embedded in a halo of mass $M_H$ was proposed by Fridman & Polyachenko (1984, §3.2 in Chapter IV). In the limit $h, M_H \to 0$, this criterion produces the marginal value $t = 0.1286$ for stellar Maclaurin disks (Kalnajs & Athanassoula 1974) and is thus equivalent to parameter $\alpha$ (Tables 1 and 2). For finite vertical thickness, stellar disks and flattened spheroids are generally found to be stable if $M_H/M_D \gtrsim 1.1$. This value is somewhat lower than the commonly obtained marginal values of $M_H/M_D \approx 2 - 3$ (see e.g. §4.5 below).

*(d) Truncated Exponential Disk/Halo Models*: Using N–body experiments, Efstathiou, Lake, & Negroponte (1982) studied the stability of truncated exponential stellar disks embedded in a variety of "halos." They found empirically a new stability criterion which was expressed as

$$\epsilon_m \equiv \frac{v_m}{(GM_D/R_S)^{1/2}} \gtrsim 1.1, \quad (3.8)$$

where $v_m$ is the maximum rotational velocity, $M_D$ is the mass, and $R_S$ is the scale length of the exponential surface density distribution in the disk. For a purely self–gravitating infinite exponential disk with no random motions (i.e. $t=1/2$), the rotation curve peaks at $R_D = 2.15006 R_S$ where $\epsilon_D = 0.62213$. Efstathiou *et al.* (1982) concluded that a hot halo component around an initially cold disk that increases $\epsilon_m$ from the value $\epsilon_D$ to more than 1.1 will provide stability to bar formation.

In what follows, we derive an approximate relation between $\alpha$ and $\epsilon_m$ and we estimate the critical value of $\alpha$ for stability implied by equation (3.8). Consider an isolated, cold, self–gravitating, exponential disk of mass $M_D$ and characteristic (maximum) rotational velocity $v_D$ in equilibrium. The virial theorem demands that

$$2T_D + W_D = 0. \quad (3.9)$$

Consider next the same disk embedded in a responsive halo and rotating with a higher velocity $v_m$. For simplicity, we assume that the halo represents an increase $M_H$ to the mass of the disk. The characteristic rotational velocity of the disk/halo system can then be written as

$$v_m^2 = v_D^2 + v_H^2, \quad (3.10)$$

where $v_H$ is the contribution of the halo. We also define the mass ratio

$$q \equiv \frac{M_D}{M_D + M_H}. \quad (3.11)$$

Combining equations (3.10) and (3.11) and using $(v_H/v_D)^2 = M_H/M_D$ we find that

$$q = \left(\frac{v_D}{v_m}\right)^2. \quad (3.12)$$

We determine next the ratio $t \equiv T/|W|$ for the original disk in the disk/halo system. Since the rotational kinetic energy of the isolated disk is $T_D \propto M_D v_D^2$ and that of the embedded disk is $T \propto M_D v_m^2$, we can write $T/T_D = (v_m/v_D)^2$. For the total gravitational potential energies we write $W_D \propto M_D^2/R_S$ and $W \propto (M_D + M_H)^2/R_S$ implying that $W_D/W = q^2$. Combining these ratios of energies with equation (3.9), we find that

$$t = \frac{1}{2}q^2\Big(\frac{v_m}{v_D}\Big)^2, \tag{3.13}$$

or, using equation (3.12), that

$$t = \frac{q}{2} = \frac{1}{2}\Big(\frac{v_D}{v_m}\Big)^2. \tag{3.14}$$

In the limit $v_m = v_D$ ($M_H = 0$), this expression reduces, as it should, to equation (3.9) and to $q = 1$. Substituting equation (3.14) in equation (3.4) and using $f=1$ (for disks) and the definition of $\epsilon_m$ we find that

$$\alpha = \frac{v_D}{2v_m} = \frac{\epsilon_D}{2\epsilon_m}. \tag{3.15}$$

We have mentioned above that Efstathiou *et al.* (1982) suggest a critical value of $\epsilon_m \approx 1.1$ for stability. Using this estimate along with $\epsilon_D = 0.62213$ in equations (3.12)–(3.15), we find that $q = 0.32$, $t = 0.16$, and $\alpha = 0.28$. Despite all the above simplifying assumptions, this result is in agreement with our critical value of $\alpha \approx 0.26$ (Table 2). [Note that using the safer value $\epsilon_m = 1.2$ instead of 1.1 we find from equation (3.15) that $\alpha = 0.26$ exactly, as well as $q = 0.27$ and $t = 0.135$.] These estimates agree reasonably well with the results given in part (b) above, the OP stability criterion, and the results given in Table 6 below for additional composite stellar disk/responsive halo models.

A stability criterion analogous to equation (3.8) but applicable to gaseous disks has not been previously established. To obtain such a criterion, we use the appropriate stability condition $\alpha \lesssim 0.34 - 0.35$ into equation (3.15). The result is $\epsilon_m \gtrsim 0.9$ for stability of gaseous disks. Using this value and equations (3.12), (3.14), we also find that $t \lesssim 0.24$ and $q \lesssim 0.48$ for stability of gaseous disk models embedded in a halo.

## 4  Differentially Rotating Centrally Condensed Models

In this section, we determine critical values of the parameter $\alpha$ for previously published, differentially rotating, centrally condensed, gaseous and stellar, numerical models. For applications, we adopt equation (3.4) as the principal expression of $\alpha$ because this equation is more transparent and straightforward to use than equation (3.6). We use equation (3.3) only in two cases of different topology where the principal expression cannot be applied because the function $f$ is unknown (§4.3 and §4.4a).

### 4.1  Bodenheimer–Ostriker (1973) Gaseous Spheroids

These differentially rotating, centrally condensed models were studied for stability by Ostriker & Bodenheimer (1973). Because of numerical difficulties, Bodenheimer & Ostriker (1973) were unable to calculate the structure of very flattened models. We have chosen the terminal models from three equilibrium sequences with polytropic and rotation indices $(n, n')$=(1.5,0), (1.5,1), and (3.0,0), the only sequences extending substantially past the Jacobi bifurcation points where $T/|W| \approx 0.14$. The model with $(n, n')$=(1.5,1) is not the last stable (marginal) model of the sequence (see Figure 2 in Ostriker & Bodenheimer 1973) and is included only for comparison purposes.

Characteristic parameters for these models are listed at the top of Table 3 along with our determination of the parameters $e$, $f$, and $\alpha$ from equations (2.1), (2.12), and (3.4), respectively. The model with $(n, n')$=(1.5,0) is confirmed to be marginally stable since $\alpha \approx 0.34$. Parameter $\alpha$ also confirms that the $(n, n')$=(3.0,0) model is nearly marginal ($\alpha \approx 0.33$) despite the fact that $T/|W| = 0.241$, a value substantially lower than the commonly quoted critical value of $T/|W| \approx 0.27$ (see also Figure 3 in Ostriker & Bodenheimer 1973). We thus have the first indication that $\alpha$ is more sensitive than $T/|W|$ as a stability indicator in gaseous models.

Using the tabulated values of $L$, $W$, and $\rho_c$ from Bodenheimer & Ostriker (1973) and our values of $\alpha$ in equations (3.6), (3.7), we have obtained estimates of the ratio $\Omega_J^2/G\rho_c$, where $\rho_c$ is the central density. The results are listed at the bottom of Table 3 for comparison purposes. Accounting for the dependence of the ratio $\Omega_J^2/G\rho_c$ on the polytropic index $n$ (seen in the last model), $\Omega_J^2$ is proportional to $\rho_c$ and independent of the rotation law (i.e. the index $n'$). Thus, it appears possible to obtain a reliable estimate of the Jeans frequency in centrally condensed, differentially rotating models by just considering the maximum density. We make use of this result in §4.3 below in order to estimate $\Omega_J$ for toroidal models.

### 4.2 Tohline–Durisen–McCollough (1985) Gaseous Spheroids

This study includes four models with $T/|W| > 0.28$. Characteristic parameters for these models are listed in Table 4 along with our determination of the parameters $e$, $f$, and $\alpha$ from equations (2.1), (2.12), and (3.4), respectively. As is also listed, only model 1 was found to be stable. Tohline *et al.* (1985) conjectured that numerical dissipation suppressed the instability. We find that $\alpha \approx 0.35$ for model 1. Therefore, model 1 appears to be marginal although $T/|W| = 0.281$. This is the second case where $\alpha$ predicts accurately stability in gaseous models.

### 4.3 Woodward–Tohline–Hachisu (1994) Gaseous Tori

This is a complete experimental study of self–gravitating tori orbiting around a central point–mass. The work expands on the results of a previous study of purely self–gravitating tori done by Tohline & Hachisu (1990). These centrally condensed, differentially rotating tori represent a very interesting class of gaseous models because purely self–gravitating systems become unstable at $T/|W|$ values as low as 0.16–0.17 while mildly self–gravitating systems are stable at $T/|W|$ values as high as 0.43–0.44 (see below). Thus, the toroidal models provide a crucial test for the validity of $\alpha$ as a stability indicator. Equation (3.4) cannot be applied to tori because of the change in topology. We return therefore to equation (3.3) and we determine approximately the ratio $\Omega/\Omega_J$ at the pressure maximum of each torus. Woodward *et al.* (1994) provide values of the ratio $t \equiv T/|W|$ and the self–gravity parameter

$$p^2 \equiv \frac{4\pi G\rho}{\Omega^2}, \qquad (4.1)$$

at the pressure maximum of each model.

To obtain an approximate estimate of $\Omega_J$ we consider first a slender, incompressible torus (cf. Goodman & Narayan 1988), a model with constant density $\rho$ and circular cross-sections. Applying Gauss's law to a cylindrical surface of radius $r$ centered around the center of the torus, we find that the potential gradient is

$$\frac{d\Phi}{dr} = 2\pi G\rho r, \qquad (4.2)$$

and thus

$$\Omega_J^2 = 2\pi G\rho. \qquad (4.3)$$

We adopt equation (4.3) for centrally condensed tori with $\rho$ as the maximum density of each model. Combining equations (3.3), (4.1), and (4.3) we find that at the pressure maximum

$$\frac{\Omega}{\Omega_J} = \frac{\sqrt{2}}{p}, \qquad (4.4)$$

and

$$\alpha = \frac{pt}{\sqrt{2}}. \qquad (4.5)$$

Model parameters from four sequences with torus/central point–mass ratios of $M_T/M_C = \infty$, 5, 1, and 0.2 are listed in Table 5 along with our determination of $\alpha$ from equation (4.5). As is also listed, Woodward *et al.* (1994) find that only model E15 is stable. Parameter $\alpha$ confirms that this is the only stable model in Table 5. The first model listed in each sequence is relatively close to the point of marginal stability. Parameter $\alpha$ confirms this result only for models E15 and E21. Notice in particular how clearly $\alpha$ distinguishes stability between models E15–E16 and between models E21–E22.

Model E8 with $T/|W| = 0.171$ and $\alpha = 0.415$ is not marginal in agreement with the more detailed study of purely self–gravitating tori by Tohline & Hachisu (1990) who find that a model with $T/|W| = 0.167$ is unstable. Model E31 with $T/|W| = 0.460$ is not marginal either. It was just the first model on the $M_T/M_C = 0.2$ sequence where the I–mode was detected with certainty. [We note that $\alpha > 1/2$ for this model implying that our approximations and equation (4.5) break down at such high values of $T/|W|$.]

Models E27–E30 (not listed in Table 5) along the $M_T/M_C = 0.2$ sequence had $T/|W| = 0.422, 0.432, 0.447$, and $0.453$. The first two models appeared to be stable while models E29 and E30 were subject to a mode of instability which Woodward *et al.* (1994) call L–mode. We find that $\alpha = 0.260, 0.319, 0.422$, and $0.487$ for models E27–E30, respectively. Therefore, parameter $\alpha$ predicts that models E27 and E28 should be stable despite the high values of $T/|W|$. Interestingly enough, no unstable mode was detected with certainty in these two models.

The above results constitute the third test that parameter $\alpha$ has passed successfully in gaseous models.

### 4.4 "Discrepant" Stellar Models

*(a) Toomre–Zang Models*: Miller (1978) studied numerically Toomre–Zang stellar disk models because linear stability analysis by Zang (1976) had indicated that nonaxisymmetric instabilities are stabilized before the axisymmetric instability with increasing velocity dispersion in such models. The Toomre–Zang models are Mestel (1963) disks with a $1/R$ surface density profile and a flat rotation curve in which only a fraction $q$ of the particles is allowed to respond to perturbations ("active" particles); particles near the center are kept frozen on their original circular orbits ("inert halo" particles). This boundary condition changes the topology of the models since the disks can no longer be regarded as simply–connected systems. Miller parameterized the initial models in terms of $q$ and the local Toomre (1964) parameter for axisymmetric stability $Q$ and defined an appropriate $t$ parameter directly comparable to the ratio $T/|W|$ used by Ostriker & Peebles (1973). He demonstrated that a model with $q=0.6$, $t=0.248$, and $Q=1.60$ was dynamically stable in clear contradiction with the OP criterion.

When $q < 1$ in the Toomre–Zang models, equation (3.4) with $f = 1$ does not apply because the disk surfaces with frozen central regions are doubly–connected. In order to obtain an estimate of the parameter $\alpha$ in such models, we relate $q$ to the ratio $\Omega/\Omega_J$ [cf. equation (3.12) in §3.4d] and, then, as in §4.3, we use equation (3.3). We think of Miller's models as active disks of mass $q$ embedded in an "inert halo" of mass $1 - q$ (see also Zang 1976), i.e. they have an equivalent "disk-to-halo" mass ratio of $M_D/M_H = q/(1 - q)$. In equilibrium, the total angular velocity $\Omega$ can be thought as the sum of two contributions, one from the self–gravity of the active disk ($\Omega_J$) and another from the gravity due to the "halo" ($\Omega_H$), i.e.

$$\Omega^2 = \Omega_J^2 + \Omega_H^2. \tag{4.6}$$

Assuming that the ratio $(\Omega_J/\Omega_H)^2$ is equal to the mass ratio $M_D/M_H$ leads to the expression

$$\frac{\Omega}{\Omega_J} = q^{-1/2}. \tag{4.7}$$

Combining equations (3.3) and (4.7) we find that

$$\alpha = tq^{1/2}, \tag{4.8}$$

which predicts that $\alpha = 0.192$ for $t=0.248$ and $q=0.6$, a value substantially smaller than the point of dynamical instability for stellar systems ($\alpha \approx 0.25$). Therefore, Miller's model with $t=0.248$ is stable according to parameter $\alpha$ and the discrepancy is resolved.

Miller (1978) also predicted by extrapolation of his results that a model with $q=0.85$, $Q=1.71$, and $t=0.282$ should be exactly marginal. We find from equation (4.8) that $\alpha =$ 0.260 for such a model, a value that implies precisely marginal stability. We thus have the first indication that $\alpha$ is more sensitive than $T/|W|$ as a stability indicator in stellar systems as well.

(b) *Spherical Models*: Miller & Smith (1980) found that stellar, purely self-gravitating systems with the structure of an $n = 3$ spherical polytrope were stable if they were forced to rotate so that $t \lesssim 0.27$ initially. Specifically, two models (T30 and T40) with initial values of $t= 0.31$ and 0.40 settled down to a final value of $t \approx 0.27$ after suffering bar instabilities. Model T30 was subject to a weak bar mode, hence it presumably lies relatively close to the critical $t$–value. We find from equation (3.4) with $f=2/3$ (spheres) that $\alpha = 0.32$ and 0.37 for models T30 and T40, respectively, in reasonable agreement with the experimental results (taking into account that most of the initial spherical models, including T30 and T40, were not in equilibrium once they were forced to rotate).

In general, adopting $f=2/3$ and exactly $\alpha=0.25$ for marginally stable spherical stellar models, we find from equation (3.4) that the critical value of $t$ for stability is $3/16(=0.1875)$ which explains the stability of all studied models with $t \leq 0.17$ (hence $\alpha \leq 0.24$) initially and in particular model RV which was an exact, high–quality equilibrium even with rotation. [Compare the values $\alpha \leq 0.24$ and $t \leq 0.17$ with the results in §3.4a and in Vandervoort (1983) who finds $t=0.1882$, corresponding to $\alpha = 0.2505$, for marginal spherical models.] This is the second case where $\alpha$ predicts accurately stability in stellar systems.

### 4.5   Additional Composite Stellar Disk/Halo Models

These differentially rotating models consist of a particle disk of mass $M_D$ embedded in a responsive particle halo of mass $M_H$ within the radius of the disk (see Shlosman & Noguchi 1993 for details). Using equation (3.4) with $f=1$ (disks), we have determined the corresponding values of $\alpha$ for four models of varying disk-to-total mass ratio $q \equiv M_D/(M_D + M_H)$. The results are listed in Table 6. As is also listed, model 2 with $q = 0.3$ ($M_H/M_D = 7/3$) and $t=0.140$ is marginal — to be precise, weakly unstable — in reasonable agreement with the OP criterion. For model 2, we find that $\alpha = 0.264$, a value that does imply marginal stability (or weak instability) as well. These results constitute the third test that parameter $\alpha$ has passed successfully in stellar models.

### 5    SUMMARY

Motivated by the results described in CKST and by a suggestion due to Vandervoort (1982, 1983), we have sought a new stability indicator for $m = 2$ (toroidal) nonaxisymmetric modes of dynamical instability in uniformly and differentially rotating, gaseous and stellar, axisymmetric systems. These modes are also known as $m = 2$ Intermediate or I–modes and bar modes in gaseous and stellar systems, respectively. In §2, we have summarized the equilibrium properties and stability characteristics of a class of uniformly rotating, self–gravitating, axisymmetric systems, the Maclaurin spheroids and disks. In §3.2, we have used these systems to formulate a new stability parameter $\alpha$ in the case of uniform rotation. We

have found that
$$\alpha \equiv \frac{T/|W|}{\Omega/\Omega_J}, \qquad (5.1)$$
where $0 \leq \alpha \leq 1/2$, $T/|W|$ is the ratio of the rotational kinetic energy to the absolute value of the gravitational potential energy, $\Omega$ is the rotation frequency, and $\Omega_J$ is the Jeans frequency introduced by self–gravity. Parameter $\alpha$ can also be expressed as a combination of the ratio $t \equiv T/|W|$ and a function $f$ that describes the topological and geometrical structure of a system, i.e.
$$\alpha = (\frac{1}{2}ft)^{1/2}. \qquad (5.2)$$
For spheroids, $f$ is a function of the meridional eccentricity only, while $f = 1$ for disks and $f = 2/3$ for spheres (see §2.1). For toroidal models, $f = p^2 t$ approximately, where $p^2 \equiv 4\pi G\rho/\Omega^2$ and the density $\rho$ and the rotation frequency $\Omega$ are evaluated at the pressure maximum of each model (see §4.3). Dynamical instability sets in at $\alpha = 0.341$–$0.354$ in gaseous systems and at $\alpha = 0.254$–$0.258$ in stellar systems (see Table 2). A comparison between the latter critical values and previously obtained criteria for stability of specific stellar models has been presented in §3.4.

The small differences in the critical values between disks and spheroids (see Table 2) indicate that parameter $\alpha$ is a reliable indicator of stability to $m = 2$ modes. The different critical values between gaseous and stellar systems ($\alpha \lesssim 0.34 - 0.35$ and $\alpha \lesssim 0.25 - 0.26$, respectively) are attributed entirely to the anisotropy of the stress tensor in stellar systems (see CKST for a detailed discussion). Besides this difference, the stable oscillatory modes that eventually produce instability appear to be similar in stellar and gaseous systems provided that the modes of gaseous systems are examined in a rotating coordinate frame so that some of the rotation is viewed as circulation analogous to that in rotating stellar models (Freeman 1966a,b,c; Hunter 1974). The transformation to a rotating frame reconciles the different definitions of $\Omega$ used in the two types of systems (see §3.1). Its effects on the stable modes of gaseous Maclaurin spheroids are illustrated in Figure 7.

Searching for the underlying physical concept, we rewrite equation (5.1) in the form given in §3.3, i.e.
$$\alpha = T_J/|W|, \qquad (5.3)$$
where
$$T_J \equiv \frac{1}{2}L\Omega_J, \qquad (5.4)$$
and $L$ is the total angular momentum. This expression of $\alpha$ clearly calls for a shift of the attention from the total rotational kinetic energy $T$ to the product $L\Omega_J$ that combines the total angular momentum with self–gravity.

The physical significance of the parameter $\alpha$ becomes transparent when equation (5.3) is written in the alternative form
$$\alpha = \frac{5f}{4}\frac{L/M}{\Omega_J a_1^2}, \qquad (5.5)$$
where $5f/4 \approx 1$ for spheroids and $f = 1$ for disks, $M$ is the mass, and $a_1$ is the equatorial radius. The term $\Omega_J a_1^2$ represents the maximum angular momentum of a circular orbit in the equatorial plane of the system. Thus, the critical values of the parameter $\alpha$ listed in Table 2 specify effectively how much the specific angular momentum $L/M$ can increase relative to $\Omega_J a_1^2$ in a Maclaurin system before dynamical instability sets in.

In §4, we have tested the validity of $\alpha$ as a stability indicator for differentially rotating, centrally condensed, numerical models that already exist in the literature. It is generally difficult to extend equations (5.1), (5.3), or (5.5) to such models because it is not obvious how weighted values of $\Omega$ and $\Omega_J$ should be estimated. We have, however, tested $\alpha$ using equation (5.2) and we find that the critical values listed in Table 2 predict accurately the onset of $m = 2$ dynamical instability in a variety of gaseous and stellar, axisymmetric models with different geometries, density distributions, and rotation laws. Thus, equation (5.2) naturally becomes the principal expression of the parameter $\alpha$ for applications. Exceptions are two topologically different models, the tori of §4.3 and the Toomre–Zang disks of §4.4a, where equation (5.2) could not be applied — the functional form of $f$ is unknown for systems with multiply–connected regions — and equation (5.1) has been used instead.

Equation (5.2) can be generalized in the case of nonaxisymmetric systems by considering the effect of nonaxisymmetry on the geometry–dependent term $f$. We shall present a generalized stability criterion in a following paper (Christodoulou, Shlosman, & Tohline 1994) and we shall test its accuracy using previously published elliptical and ellipsoidal equilibrium models.

## Acknowledgments

We are grateful to J. Ostriker for helpful discussions and suggestions. We thank C. Hunter, A. Toomre, and P. Vandervoort for useful correspondence. We also thank C. Heller for recalculating some of the values in Table 6 and J. Woodward for sending us some of his results. IS is grateful to Gauss Foundation for support and to K. Fricke, Director of Universitäts-Sternwarte Göttingen, for hospitality during a stay in which much of this work has been accomplished. This work was supported in part by NASA grants NAGW–1510, NAGW–2447, NAGW–2376, and NAGW–3839, by NSF grant AST–9008166, and by grants from the San Diego Supercomputer Center, the National Center for Supercomputing Applications, and the University of Kentucky Center for Computational Studies.

TABLE 1
Points of Dynamical Bar–Forming Instability

|  | Disk | Spheroid | Variation (%) |
|---|---|---|---|
| Gaseous | $t$=0.25000 | $t$=0.27383 | $|\Delta t|/t_{Sph}$=8.70 |
|  | $\chi$=0.70711 | $\chi$=0.80306 | $|\Delta\chi|/\chi_{Sph}$=11.9 |
|  |  | $e$=0.95289 |  |
|  |  | $f$=0.84922 |  |
| Stellar | $t$=0.12860 | $t$=0.17114 | $|\Delta t|/t_{Sph}$=24.9 |
|  | $\chi$=0.50715 | $\chi$=0.66212 | $|\Delta\chi|/\chi_{Sph}$=23.4 |
|  |  | $e$=0.86362 |  |
|  |  | $f$=0.78076 |  |

TABLE 2
Critical Values of the Stability Parameter $\alpha$

|  | Disk | Spheroid | $|\Delta\alpha|/\alpha_{Sph}$ (%) |
|---|---|---|---|
| Gaseous | 0.35355 | 0.34098 | 3.69 |
| Stellar | 0.25357 | 0.25847 | 1.90 |

TABLE 3
Bodenheimer–Ostriker Gaseous Spheroids

| $(n,n')$ | $T/|W|$ | $a_1/a_3$ | $e$ | $f$ | $\alpha$ |
|---|---|---|---|---|---|
| (1.5,0) | 0.264 | 3.78 | 0.964 | 0.864 | 0.338 |
| (1.5,1) | 0.219 | 4.38 | 0.974 | 0.881 | 0.311 |
| (3.0,0) | 0.241 | 4.68 | 0.977 | 0.887 | 0.327 |
| $(n,n')$ | $L$ | $|W|$ | $\alpha$ | $\rho_c$ | $\Omega_J^2/G\rho_c$ |
| (1.5,0) | 0.340 | 1.15 | 0.338 | 5.24 | 1.0 |
| (1.5,1) | 0.256 | 1.67 | 0.311 | 16.8 | 1.0 |
| (3.0,0) | 0.252 | 1.85 | 0.327 | 33.1 | 0.7 |

TABLE 4
TOHLINE–DURISEN–MCCOLLOUGH
GASEOUS SPHEROIDS

| Model | $T/|W|$ | $a_1/a_3$ | Stability | $e$ | $f$ | $\alpha$ |
|---|---|---|---|---|---|---|
| 1 | 0.281 | 4.23 | Stable | 0.972 | 0.877 | 0.351 |
| 2 | 0.301 | 4.88 | Unstable | 0.979 | 0.891 | 0.366 |
| 3 | 0.327 | 5.96 | Unstable | 0.986 | 0.908 | 0.385 |
| 4 | 0.352 | 7.29 | Unstable | 0.991 | 0.924 | 0.403 |

TABLE 5
WOODWARD–TOHLINE–HACHISU
GASEOUS TORI

| Model | $M_T/M_C$ | $T/|W|$ | $p^2$ | Stability | $\alpha$ |
|---|---|---|---|---|---|
| E8 | $\infty$ | 0.171 | 11.8 | Unstable | 0.415 |
| E15 | 5 | 0.217 | 3.85 | Stable | 0.301 |
| E16 | 5 | 0.238 | 5.37 | Unstable | 0.390 |
| E21 | 1 | 0.342 | 2.14 | Unstable | 0.354 |
| E22 | 1 | 0.356 | 3.06 | Unstable | 0.440 |
| E31 | 0.2 | 0.460 | 3.17 | Unstable | 0.579 |

TABLE 6
SHLOSMAN–NOGUCHI
STELLAR DISK/HALO MODELS

| Model | $q$ | $M_H/M_D$ | $T/|W|$ | Stability | $\alpha$ |
|---|---|---|---|---|---|
| 1 | 0.2 | 4/1 | 0.083 | Stable | 0.204 |
| 2 | 0.3 | 7/3 | 0.140 | Marginal | 0.264 |
| 3 | 0.5 | 1/1 | 0.223 | Unstable | 0.334 |
| 4 | 1.0 | 0 | 0.320 | Unstable | 0.400 |